\documentclass[]{./elsart}

\def\aap{\ {A\&A}\ }

\def\aj{\ {AJ}\ }

\def\apj{\ {ApJ}\ }
\def\apjl{\ {ApJL}\ }
\def\apjs{\ {ApJS}\ }

\def\mnras{\ {MNRAS}\ }
\def\nat{\ {Nat}\ }

\newcommand{\msun}{\mbox{${\rm M}_\odot$}}

\newcommand{\kms}{\mbox{${\rm km~s}^{-1}$}}

\def\apgt{\ {\raise-.5ex\hbox{$\buildrel>\over\sim$}}\ }
\def\aplt{\ {\raise-.5ex\hbox{$\buildrel<\over\sim$}}\ }

\input ./psfig.sty

\begin{document}

\begin{frontmatter}

\title{The gravitational wave signature of young and dense star clusters}

\author{Simon F. Portegies Zwart}

\address{Astronomical Institute `Anton Pannekoek', 
       University of Amsterdam, Kruislaan 403, the Netherlands and
       Institute for Computer Science, University of Amsterdam, Kruislaan 403 
}

\begin{abstract}

Young star clusters are often born with such high stellar densities
that stellar collisions play an important role in their further
evolution. In such environments the same star may participate in
several tens to hundreds of collisions ultimately leading to the
collapse of the star to a black hole of intermediate mass. At later
time the black hole may acquire a companion star by tidal capture or
by dynamical --3-body-- capture. When this companion star evolves it
will fill its Roche-lobe and transfer mass to its accompanying black
hole. This then leads to a bright phase of X-ray emission but the
binary may also be visible as a source of gravitational wave radiation
in the LISA band. If the star captured by the intermediate mass black
hole is relatively low mass ($\aplt 2$\,\msun) the binary will be
visible as a bright source in gravitational waves and X-rays for the
entire lifetime of the binary.  The majority of compact binaries which
formed from the population of primordial binaries in young and dense
star clusters do not lead to detectable gravitational wave
sources. 
\end{abstract}

\begin{keyword}
black holes \sep compact objects \sep simulation \sep star clusters
\sep gravitational waves
\PACS code \sep code \PACS
\end{keyword}
\end{frontmatter}

\section{Introduction}\label{Sect:introduction}

Dense and young star clusters are rather rare in our Galaxy.  The six
such star clusters that are currently known are: Arches
\cite{2002ApJ...581..258F}, IRS13 \cite{2004A&A...423..155M} and IRS16
\cite{2003ApJ...593..352P,2004ApJ...602..760E}, NGC3603
\cite{1999A&A...352L..69B}, Quintuplet \cite{1999ApJ...514..202F} and
Westerlund~1 \cite{2000ApJ...533L..17V}.  The two IRS sources,
however, can hardly be called a cluster as they look more like the
debris of disrupted star cluster
\cite{2000ApJ...545..301K,2003ApJ...596..314M}.  Young and dense star
clusters (hereafter YDC, sometimes pronounced as YoDeC) are
characterized by a stellar population which is younger than about
10\,Myr.

The fact that these clusters are young means that stars of all masses
are still present, offering critical insights into the stellar initial
mass function and cluster structural properties at formation.  The
term ``dense'' means that dynamical evolution and physical collisional
processes can operate fast enough to compete with and even overwhelm
stellar evolutionary timescales; dense stellar systems are places
where wholly new stellar evolution channels can occur, allowing the
formation of stellar species completely inaccessible by standard
stellar and binary evolutionary pathways.

There are no examples of the older ($\apgt10$\,Myr) siblings of these
clusters in the Milky Way Galaxy. For the Arches and Quintuplet this
may not be so surprising, as they dissolve in the tidal field of the
Galaxy in at most a few tens of million years
\cite{2000ApJ...545..301K,2002ApJ...565..265P}, and even if they
survive longer they will become hard to detect against the dense
background stellar population \cite{2001ApJ...546L.101P}. For the two
rather isolated clusters, NGC\,3603 and Westerlund~1, it is rather
curious that there are no examples of their 10 to 100\,Myr descendants
in the Milky Way.

The other characteristic of YDCs is the high density. The few clusters
near the Galactic center must have a high density, as otherwise they
would be disrupted easily by the strong tidal field. The clusters
further out however, do not have this constraint and it is interesting
to note that clearly these clusters ware born without much knowledge
of the local potential of the Galaxy, i.e: they behave as isolated
clusters.

In this paper, we discuss the formation and evolution of gravitational
wave sources in the form of compact binaries in YDCs, but for
convenience we extend our definition to star clusters of 100\,Myr.
The basis of this study is the star cluster MGG11
\cite{2003ApJ...596..240M} at about 200\,pc from the nucleus of the
starburst galaxy M82 at a distance of about 3.6\,Mpc.

The star cluster MGG-11 has an age of about 7--12\,Myr
\cite{2003ApJ...596..240M}, a line-of-sight velocity dispersion of
$\sigma_r = 11.4\pm0.8\,\kms$ and a projected half-light radii, $r =
1.2$\,pc. The cluster mass then totals $\sim 3.5 \times 10^5
\msun$. The cluster mass function seems to be deficient of stars below
about 1\,\msun, but follow a Salpeter slope for the higher masses.

We simulate this cluster by integrating the equations of motion for
all stars and follow the evolution of stars and binaries in the
cluster using the {\tt starlab}\footnote{{\tt
http://www.ids.ias.edu/$\sim$starlab/}} \cite{2001MNRAS.321..199P}
direct N-body package.  We start our simulations with 144179 stars,
13107 of which are in binaries totaling a binary fraction of about
10\%.  The binary parameters ware selected as follows: first we chose
a random binding energy between a minimum of 10kT and a maximum, and
we chose the orbital eccentricity from the thermal distribution. The
maximum binding energy was selected such that the distance at
pericenter exceeded four times the radius of the primary star. For the
initial density profile we adopt a King model with $W_0=12$
\cite{1966AJ.....71...64K}. We ignored an external tidal field, but
stars are removed from the simulation if they are more than 10 tidal
radii (about 80\,pc) away from the density center of the cluster.

\section{Global cluster evolution}
\subsection{Early core collapse and the growth of an 
	intermediate mass black hole}

\begin{figure}
\psfig{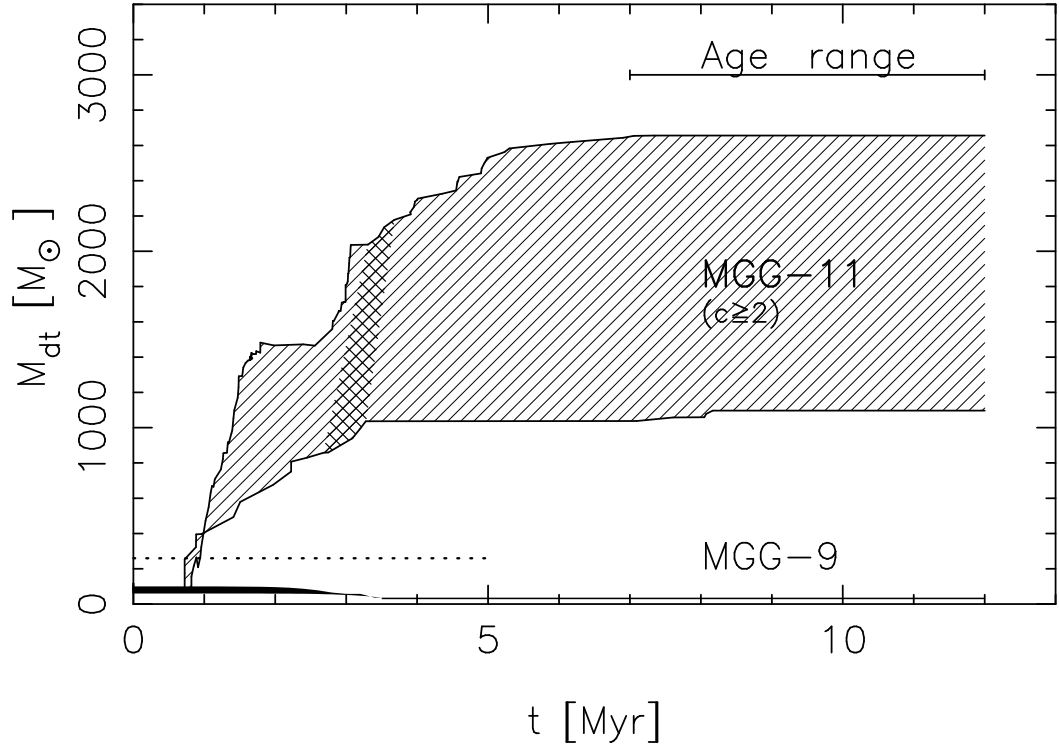} 
\caption[]{ The shaded area indicates the range in mass for the
designated target $M_{\rm dt}$ as a function of time for simulations
which resemble the star cluster MGG\,11 (see
\S\,\ref{Sect:introduction}).  This area is calculated with a large
variety in initial conditions, and computed with two independently
developed N-body codes {\tt NBODY4} and {\tt starlab} (see
\cite{2004Natur.428..724P} for details).  The hashed area indicates
the moment in time where the collision runaway typically collapses to
a black hole of intermediate mass. The age range for the star cluster
MGG\,11 is given by the upper horizontal bar between 7\,Myr and
12\,Myr. It's neighboring cluster, MGG-9, is not expected to have
experienced a collision runaway and the evolution of the mass of the
most massive star in this cluster is indicated by the lower solid
line. The horizontal dotted curve indicates the mass of the most
massive star initially present in the calculations.  }
\label{fig:Mbh_all}
\end{figure}

Driven by the massive stars, the cluster experiences an early core
collapse \cite{2004Natur.428..724P,2004ApJ...604..632G}, which leads
to subsequent collisions between massive stars. The result of this is
the growth of one of the initially most massive stars through repeated
collisions, to a total mass of about 1000-3000\,\msun.  The evolution
of the mass of this {\em designated target} is illustrated in
figure\,\ref{fig:Mbh_all}, where we show the range in masses acquired
by the star that experiences repeated collisions. In the end the
massive star may collapse to an intermediate mass black hole
\cite{2004Natur.428..724P}\footnote{Several alternative theoretical
models exist for producing black holes of $\sim10^2-10^4 M_\odot$;
Portegies Zwart \& McMillan 2002; Miller \& Hamilton 2002; Madau \&
Rees 2001).\nocite{2002ApJ...576..899P} \nocite{2002MNRAS.330..232M}
\nocite{2001ApJ...551L..27M}.}.

In a recent model Hopman et al.\, \cite{2004ApJ...604L.101H} proposed
that an intermediate mass black hole can capture a companion star in a
tight orbit.  Further tidal interaction between the black hole and the
captured star then circularize the orbit.  They further assume that
the captured star is on the main-sequence, but the same argument can
be made for evolved stars.  By the time the orbit has been fully
circularized the captured star under-fills its Roche-lobe only
slightly. During the remainder of the main-sequence lifetime of the
captured star, it grows in size by about a factor of two and
gravitational wave radiation reduces the orbital
separation. Ultimately the star fills its Roche lobe and starts to
transfer mass to the intermediate mass black hole.

We do not further consider the details of tidal capture, nor do we
assess the probability of forming the type of binary systems discussed
in this paper, as we mainly concentrate on the phase in the lifetime
of the intermediate mass black hole at which it is visible as a source
for gravitational waves.

\subsection{The further evolution of the intermediate mass black hole}

After tidal capture we evolved the binary through various stages using
the binary evolution code of Eggleton (Pols et
al. 1995\nocite{1995MNRAS.274..964P} and references therein), assuming
a population I chemical composition (Y = 0.98, Z = 0.02),
mixing-length parameter of $\alpha = 2.0$ and with convective
overshooting constant $\delta_{\rm ov} = 0.12$ (Pols et
al. 1998)\nocite{1998MNRAS.298..525P}.  Results on such evolution are
published by Portegies Zwart et al \cite{2004astro.ph..8402Z}.  They
calculated the orbital evolution of the binary systems, as they are
affected by the emission of gravitational waves (Landau \& Lifshitz,
1958),\nocite{1958.book.....L} Roche-lobe overflow and by
non-conservative mass transfer or by mass loss via a wind (Soberman et
al. 1997) \cite{1997A&A...327..620S}.  Matter was assumed to leaves
the system carrying the specific angular momentum of the donor. During
mass-transfer the black hole was assumed to accrete matter up to its
Eddington limit (see e.g.\, King 2000)\nocite{2000MNRAS.312L..39K}.

\begin{figure}
\psfig{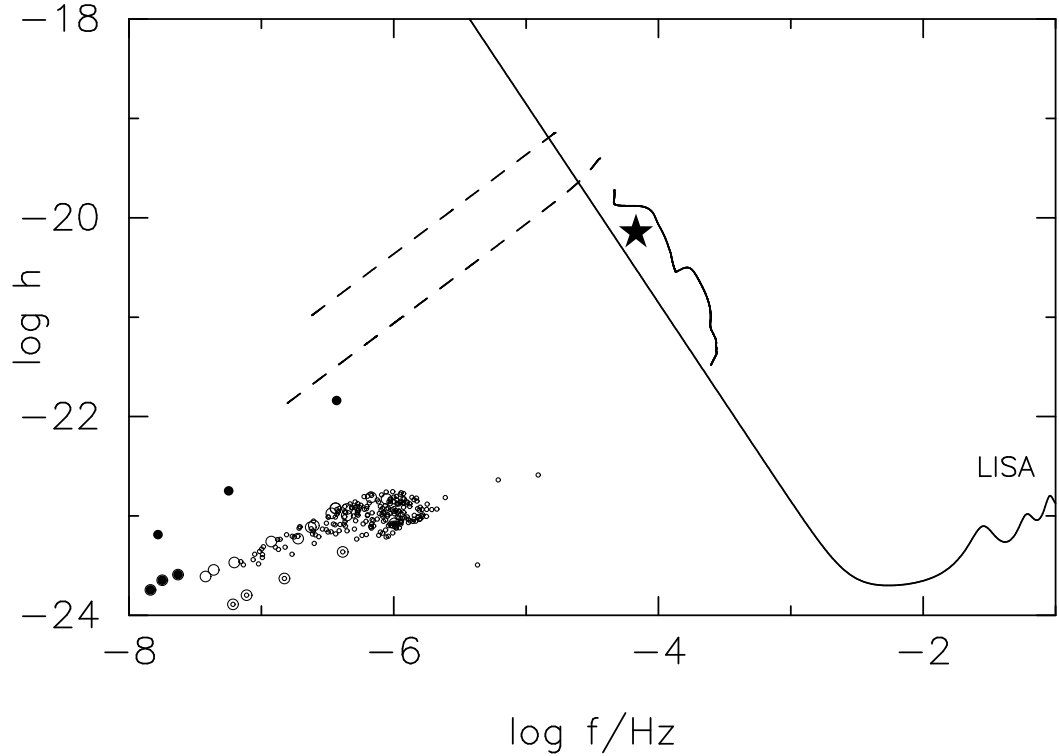} 
\caption[]{ A selected sample of binaries in the gravitational wave
strain and frequency domain. The LISA noise curves is over plotted as
a reference.  For all sources we assumed a distance to the source of
1kpc.  The two diagonal dashed curves indicate the range of parameters
(wave strain and frequency) between which a 5\,\msun\, (lower) and
15\,\msun\, (upper) main-sequence star which transfers mass to a
1000\,\msun\, black hole is visible. The solid curve gives the
evolution of the 2\,\msun\, main-sequence star which transfers mass to
a 1000\,\msun\, black hole. The other symbols indicate the position of
compact binaries from the in this paper described N-body simulation of
MGG-11 at an age of 100\,Myr. The large $\star$ gives the position of
the $\sim 1130$\,\msun\, intermediate mass black hole, which happens
to have acquired a 1.4\,\msun\, main-sequence companion star in a
tight orbit.  Each binary with two compact objects in this simulation
(at an age of 100\,Myr) is indicated with a single symbol or with a
combination of symbols. Each symbols stands for either a stellar mass
black hole (bullet), a neutron star (large circle) or a white dwarf
(small circle). Some of the bullets to the left (binaries with two
black holes) will eventually evolve through the LISA detection limit
and become visible as gravitational wave sources.  }
\label{fig:GWR_cluster}
\end{figure}

The binary emits gravitational waves, and as long as the stars are far
apart we can assume that they behave as point masses. Note, however,
that the donor star fills its Roche-lobe and can hardly be described
as a point mass.  The amplitude of the gravitational wave signal is
given by (Evens et al.\, 1987)
\begin{equation}\label{eq:h}
  h \simeq 6 \times 10^{-23} \left( \frac{\mathcal{M}}{\msun} \right)^{5/3}
\! \! \! \left( \frac{P_{\rm orb}}{\rm 1\, {\rm days}} \right)^{-2/3} \!
\! \!\left( \frac{d}{\rm 1 {\rm kpc}} \right)^{-1}.
\end{equation}
Here $\mathcal{M}$ is the chirp mass.
The gravitational wave frequency is twice the orbital frequency.  The
resulting gravitational wave signal is presented in
Fig.\,\ref{fig:GWR_cluster}, for binaries which started Roche-lobe
overflow on the zero-age main-sequence. Binaries which do not stars
RLOF during the main-sequence phase of the donor will not become
detectable gravitational wave sources.

At the onset of Roche-lobe overflow the wave strain is $\log h \apgt
-19.2$ for a 1000\,\msun\, black hole with a $\apgt 5$\,\msun\,
Roche-lobe filling main-sequence companion at a distance of 1kpc.
While mass transfer proceeds the binary system becomes wider,
resulting in a increase of the orbital period and consequently in a
decrease of the wave strain. A binary at a distance of $\sim 1$\,kpc
is then visible for about 1/10$^{th}$ of its main-sequence lifetime,
which, for an $\apgt 5$\,\msun\, donor is at most a few million years.

Figure\,\ref{fig:GWR_cluster} shows the expectation of the LISA noise
curve (diagonal solid curve from top left to bottom right. Sources to
the right and above this curve are expected to be detectable by the
LISA satellite. Main-sequence stars of 5\,\msun\, and 15\,\msun\,
which ware born in Roche lobe contact to a 1000\,\msun\, black hole
are visible between the two diagonal dashed curves, assuming a
distance of 1\,kpc. Note that the star cluster Westerlund\,1 is at a
distance of 1.1\,kpc.

The $\aplt 10$\,\msun\, donors turn into a white dwarfs after their
entire hydrogen envelope has been transferred to the black hole. After
that, the emission of gravitational waves brings the two stars back
into the relatively high frequency regime and it becomes detectable
again for the {\em LISA} antennae. This process, however takes far
longer than a Hubble time, unless the orbit has an eccentricity $e >
0.97$. Such high eccentricities, after the phase of mass transfer, can
only be achieved in the cases where the donor collapses to a compact
object in a supernova explosion, or if the binary is perturbed by
external influences.  Note that the kick velocity due to a possible
asymmetry in the supernova in which the neutron star forms is not
sufficient to unbind the binary system.  The supernova may induce a
high eccentricity, leading to a relatively quick merger due to the
emission of gravitational waves. Such a post-supernova system can be
seem by LISA over a much larger distance, enhancing the available
volume within which such event can occur.  We conclude that the here
discussed class of binaries with relatively massive donors $\apgt
5$\,\msun\, in a state of Roche-lobe contact are probably not
important sources of gravitational waves, because this phase is short
lived and the gravitational wave radiation emitted is relatively
weak. If the donor collapses to a neutron star, however, the binary
can be seen to a much larger distance and this may result in an
appreciable detection rate.

The binary in which a 2\,\msun\, main-sequence star starts to transfer
mass to a 1000\,\msun\, black hole, however, remains visible as a
bright source of gravitational waves for its entire lifetime (see
Figure\,\ref{fig:GWR_cluster}). The reason for this striking result is
the curious evolution of the donor.  During RLOF, the hydrogen content
in the star is continuously fully mixed, and as a consequence the star
remains rather small in size, causing the mass transfer rate to
remains roughly constant.  By the time the hydrogen fraction drops
below $\sim 0.66$, the central temperature becomes so low that
thermonuclear fusion stops: this happens at about $t=630$\,Myr.  At
that point the system is detached for about 8\,Myr before it undergoes
another phase of mass transfer until $t=1.2$\,Gyr, at which point the
donor turns into a 0.011\,\msun\, (about 10 $m_{\rm Jup}$) brown
dwarf. The brown dwarf ultimately merges with the black hole due to
the emission of gravitational-waves \cite{2004astro.ph..8402Z}.

This binary remains visible in the LISA frequency regime for its
entire lifetime, only the gravitational wave strain drops as the donor
is slowly consumed by the intermediate mass black hole.  Mass transfer
in this evolutionary stage is rather slow causing the X-ray source to
be transient, being bright for a few days every other month. We argue
that such a transient could result in an interesting synchronous
detection of X-rays and gravitational waves
\cite{2004astro.ph..8402Z}.

The solid curve in the middle of the diagram gives the gravitational
wave radiation emitted by a 2\,\msun\, star transferring mass to a
1000\,\msun\, black hole. This binary has a rather peculiar evolution
\cite{2004astro.ph..8402Z}.  Such binaries may remain visible in
X-rays and as a source of gravitational waves for their entire
main-sequence lifetime.  The gravitational waves, however will only be
visible by the LISA antennae do a distance of at most a few kpc.

The other symbols in figure\,\ref{fig:GWR_cluster} give the location of
the compact binaries in the frequency--strain diagram for the
simulated cluster at an age of 100\,Myr.  
The large $\star$\, identifies the
location of an intermediate mass black hole formed in the simulations,
which, at that moment had a mass of about 1130\,\msun\, and had a
1.4\,\msun\, main-sequence star companion in a 0.6\,day orbit. This
binary was captured by a three-body process.

Stellar-mass black holes, neutron stars and white dwarfs are
represented in figure\,\ref{fig:GWR_cluster} by bullets, large circles
and small circles, respectively.  All objects are binaries, and a
single large circle therefore represents a binary consisting of two
neutron stars, of which there are a couple. The few large circles with
an inner smaller circle (near $\log f \sim -7$ and $\log h \simeq
-23.7$ are binaries with a neutron star and a white dwarf.  The
majority of objects contain black holes or white dwarfs; binaries with
neutron stars are rare because of the large kick velocities they
receive upon birth.

\section{Summary}

Young and dense star clusters are promising sources for gravitational
wave radiation. Especially when they get older than $\apgt 50$\,Myr as
white dwarf binaries start to form. But also at younger age the
presence of an intermediate mass black hole may provide an important
source of gravitational waves.

An interesting possibility is provided by a $\sim 2$\,\msun\, donor
which starts to transfer mass to an intermediate mass black hole at
birth. Such a binary may be visible as a bright transient X-ray source
and simultaneously as gravitational wave source in the {\em LISA} band
to a distance of a few kpc.

\section*{Acknowledgments}
I am grateful to Steve McMillan and Gijs Nelemans for numerous
discussions, and in particular to Jun for the excessive use of his
GRAPE-6, on which these simulations are performed. Additional
simulations are performed at the GRAPE-6 systems at Drexel University
and the MoDeStA computer at the University of Amsterdam.  This work
was supported by NASA ATP, the Royal Netherlands Academy of Sciences
(KNAW), the Dutch organization of Science (NWO), and by the
Netherlands Research School for Astronomy (NOVA).


\begin{thebibliography}{26}
\expandafter\ifx\csname natexlab\endcsname\relax\def\natexlab#1{#1}\fi
\expandafter\ifx\csname url\endcsname\relax
  \def\url#1{\texttt{#1}}\fi
\expandafter\ifx\csname urlprefix\endcsname\relax\def\urlprefix{URL }\fi

\bibitem[{{Brandl} et~al.(1999)}]{1999A&A...352L..69B}
{Brandl}, B., {Brandner}, W., {Eisenhauer}, F., {Moffat}, A. F.~J., {Palla},
  F., {Zinnecker}, H., Dec. 1999. Low-mass stars in the massive h ii region ngc
  3603 deep nir imaging with antu/isaac. \aap 352, L69--L72.

\bibitem[{{Eckart} et~al.(2004)}]{2004ApJ...602..760E}
{Eckart}, A., {Moultaka}, J., {Viehmann}, T., {Straubmeier}, C., {Mouawad}, N.,
  Feb. 2004. {Young Stars at the Center of the Milky Way?} \apj 602, 760--769.

\bibitem[{{Figer} et~al.(1999)}]{1999ApJ...514..202F}
{Figer}, D.~F., {McLean}, I.~S., {Morris}, M., Mar. 1999. Massive stars in the
  quintuplet cluster. \apj 514, 202--220.

\bibitem[{{Figer} et~al.(2002)}]{2002ApJ...581..258F}
{Figer}, D.~F., {Najarro}, F., {Gilmore}, D., {Morris}, M., {Kim}, S.~S.,
  {Serabyn}, E., {McLean}, I.~S., {Gilbert}, A.~M., {Graham}, J.~R., {Larkin},
  J.~E., {Levenson}, N.~A., {Teplitz}, H.~I., Dec. 2002. {Massive Stars in the
  Arches Cluster}. \apj 581, 258--275.

\bibitem[{{G{\" u}rkan} et~al.(2004)}]{2004ApJ...604..632G}
{G{\" u}rkan}, M.~A., {Freitag}, M., {Rasio}, F.~A., Apr. 2004. {Formation of
  Massive Black Holes in Dense Star Clusters. I. Mass Segregation and Core
  Collapse}. \apj 604, 632--652.

\bibitem[{{Hopman} et~al.(2004)}]{2004ApJ...604L.101H}
{Hopman}, C., {Portegies Zwart}, S.~F., {Alexander}, T., Apr. 2004.
  {Ultraluminous X-Ray Sources as Intermediate-Mass Black Holes Fed by Tidally
  Captured Stars}. \apjl 604, L101--L104.

\bibitem[{{Kim} et~al.(2000)}]{2000ApJ...545..301K}
{Kim}, S.~S., {Figer}, D.~F., {Lee}, H.~M., {Morris}, M., Dec. 2000. ``n-body
  simulations of compact young clusters near the galactic center''. \apj 545,
  301--308.

\bibitem[{{King}(2000)}]{2000MNRAS.312L..39K}
{King}, A.~R., Mar. 2000. {Black hole transients and the Eddington limit}.
  \mnras 312, L39--L41.

\bibitem[{{King}(1966)}]{1966AJ.....71...64K}
{King}, I.~R., Feb. 1966. The structure of star clusters. iii. some simple
  dvriamical models. \aj 71, 64.

\bibitem[{{Landau} and {Lifshitz}(1958)}]{1958.book.....L}
{Landau}, L.~D., {Lifshitz}, M., 1958. {The Classical Theory of Fields}.
  Pergamon Press, Oxford, London, New York, Paris.

\bibitem[{{Madau} and {Rees}(2001)}]{2001ApJ...551L..27M}
{Madau}, P., {Rees}, M.~J., Apr. 2001. Massive black holes as population iii
  remnants. \apjl 551, L27--L30.

\bibitem[{{Maillard} et~al.(2004)}]{2004A&A...423..155M}
{Maillard}, J.~P., {Paumard}, T., {Stolovy}, S.~R., {Rigaut}, F., Aug. 2004.
  {The nature of the Galactic Center source IRS 13 revealed by high spatial
  resolution in the infrared}. \aap 423, 155--167.

\bibitem[{{McCrady} et~al.(2003)}]{2003ApJ...596..240M}
{McCrady}, N., {Gilbert}, A.~M., {Graham}, J.~R., Oct. 2003. {Kinematic Masses
  of Super-Star Clusters in M82 from High-Resolution Near-Infrared
  Spectroscopy}. \apj 596, 240--252.

\bibitem[{{McMillan} and {Portegies Zwart}(2003)}]{2003ApJ...596..314M}
{McMillan}, S.~L.~W., {Portegies Zwart}, S.~F., Oct. 2003. {The Fate of Star
  Clusters near the Galactic Center. I. Analytic Considerations}. \apj 596,
  314--322.

\bibitem[{{Miller} and {Hamilton}(2002)}]{2002MNRAS.330..232M}
{Miller}, M.~C., {Hamilton}, D.~P., Feb. 2002. {Production of intermediate-mass
  black holes in globular clusters}. \mnras 330, 232--240.

\bibitem[{{Pols} et~al.(1998)}]{1998MNRAS.298..525P}
{Pols}, O.~R., {Schroder}, K., {Hurley}, J.~R., {Tout}, C.~A., {Eggleton},
  P.~P., Aug. 1998. {Stellar evolution models for Z = 0.0001 to 0.03}. \mnras
  298, 525--536.

\bibitem[{{Pols} et~al.(1995)}]{1995MNRAS.274..964P}
{Pols}, O.~R., {Tout}, C.~A., {Eggleton}, P.~P., {Han}, Z., Jun. 1995.
  {Approximate input physics for stellar modelling}. \mnras 274, 964--974.

\bibitem[{{Portegies Zwart} et~al.(2004)}]{2004Natur.428..724P}
{Portegies Zwart}, S.~F., {Baumgardt}, H., {Hut}, P., {Makino}, J., {McMillan},
  S.~L.~W., Apr. 2004. {Formation of massive black holes through runaway
  collisions in dense young star clusters}. \nat 428, 724--726.

\bibitem[{{Portegies Zwart} et~al.(2001{\natexlab{a}})}]{2001ApJ...546L.101P}
{Portegies Zwart}, S.~F., {Makino}, J., {McMillan}, S.~L.~W., {Hut}, P., Jan.
  2001{\natexlab{a}}. {How Many Young Star Clusters Exist in the Galactic
  Center?} \apjl 546, L101--L104.

\bibitem[{{Portegies Zwart} et~al.(2002)}]{2002ApJ...565..265P}
{Portegies Zwart}, S.~F., {Makino}, J., {McMillan}, S.~L.~W., {Hut}, P., Jan.
  2002. {The Lives and Deaths of Star Clusters near the Galactic Center}. \apj
  565, 265--279.

\bibitem[{{Portegies Zwart} and {McMillan}(2002)}]{2002ApJ...576..899P}
{Portegies Zwart}, S.~F., {McMillan}, S.~L.~W., Sep. 2002. {The Runaway Growth
  of Intermediate-Mass Black Holes in Dense Star Clusters}. \apj 576, 899--907.

\bibitem[{{Portegies Zwart} et~al.(2003)}]{2003ApJ...593..352P}
{Portegies Zwart}, S.~F., {McMillan}, S.~L.~W., {Gerhard}, O., Aug. 2003. {The
  Origin of IRS 16: Dynamically Driven In-Spiral of a Dense Star Cluster to the
  Galactic Center?} \apj 593, 352--357.

\bibitem[{{Portegies Zwart} et~al.(2001{\natexlab{b}})}]{2001MNRAS.321..199P}
{Portegies Zwart}, S.~F., {McMillan}, S.~L.~W., {Hut}, P., {Makino}, J., Feb.
  2001{\natexlab{b}}. {Star cluster ecology - IV. Dissection of an open star
  cluster: photometry}. \mnras 321, 199--226.

\bibitem[{{Portegies Zwart} et~al.(2004)}]{2004astro.ph..8402Z}
{Portegies Zwart}, S.~P., {Dewi}, J., {Maccarone}, T., Aug. 2004. {Intermediate Mass
  Black Holes in Accreting Binaries: Formation, Evolution and Observational
  Appearance}. ArXiv Astrophysics e-prints.

\bibitem[{{Soberman} et~al.(1997)}]{1997A&A...327..620S}
{Soberman}, G.~E., {Phinney}, E.~S., {van den Heuvel}, E.~P.~J., Nov. 1997.
  {Stability criteria for mass transfer in binary stellar evolution.} \aap 327,
  620--635.

\bibitem[{{Vrba} et~al.(2000)}]{2000ApJ...533L..17V}
{Vrba}, F.~J., {Henden}, A.~A., {Luginbuhl}, C.~B., {Guetter}, H.~H.,
  {Hartmann}, D.~H., {Klose}, S., Apr. 2000. ``the discovery of an embedded
  cluster of high-mass stars near sgr 1900+14 ``. \apjl 533, L17--L20.

\end{thebibliography}
\end{document}